# Quantitative structural and textural assessment of laminar pyrocarbons through Raman spectroscopy, electron diffraction and few other techniques

**Jean-Marie Vallerot, Xavier Bourrat(*), Arnaud Mouchon and Georges Chollon**
University of Bordeaux 1, Laboratoire des composites Thermostructuraux, 3 allée de La Boétie, F-33 600 Pessac, France
(*) Present address : ISTO – CNRS-Université d'Orléans, UMR 6113 – BP 6759, 45067 Orléans cedex 2, France

## Abstract

In pyrocarbon materials, the width of the Raman D band ($FWHM_D$) is very sensitive to low energy structural defects (e.g., disorientations of the graphene layers). The correlation between the two parameters, $FWHM_D$ and OA (as derived from selected area electron diffraction: SAED), has allowed to differentiate various pyrocarbons unambiguously. Furthermore, the optical properties of pyrocarbons, i.e., the extinction angle, the optical phase shift and the ordinary and extraordinary reflectance, have been accurately determined at 550 nm by means of the extinction curves method. These results are completed by in-plane and out-of-plane dielectric constant measurements by angular resolved EELS. Moreover, the hybridization degree of the carbon atoms has been assessed by the same technique. About 80% of the carbon atoms of the pyrocarbons have a $sp^2$ hybridization. The lack of pure $sp^2$ carbon atoms, as compared to graphite, might be explained by the presence of $sp^3$-like line defects.

## 1. Introduction

Rough laminar pyrocarbon (RL) has been extensively studied through the past decades due to a number of successful applications. Most of the commercial C/C composite matrices, processed by chemical vapour infiltration (CVI), indeed consist of RL pyrocarbon. This low temperature pyrolytic carbon matrix exhibits the best range of properties (ablation, tribological properties, Young's modulus, thermal diffusivity, density) among all its low temperature counterparts. Lieberman and Pierson [1] named it in the early seventies, referring to the aspect of its Maltese-cross when observed by polarized optical microscopy. The development of new densification methods of felts, like pulsed-CVI or thermal gradient CVI, led to another attractive pyrocarbon called regenerative laminar pyrocarbon (ReL) [2]. ReL indeed shares common characteristics with its famous predecessor. Like RL, ReL is a strongly anisotropic, graphitizable and high modulus pyrocarbon. However, it failed to dismiss RL because of its poorer properties, especially for tribological applications.

Two methods were developed aiming at an objective structural control of pyrocarbon matrices. First, the measurement of the extinction angle Ae was introduced by Diefendorf and Tokarsky [3]. This technique is very fast and easy as it only requires an optical microscope for metallography with two polarizers. The second method is the assessment of the orientation angle OA, measured from the selected area electronic diffraction (SAED), introduced by Bourrat et al. [4]. OA is the full width at half maximum (FWHM) of the Gaussian distribution of orientation of the graphene planes around the anisotropy plane. The two methods were linearly related [4], indicating they both quantify adequately the anisotropy[1] of the matrix. These characterization tools were extensively applied by authors who wished to understand the growth mechanism of low temperature pyrocarbons under various deposition conditions. New classifications were obtained in this way, with other designations related to the anisotropy of the material, like those proposed by Reznik and Hüttinger [5] or Le Poche et al. [6].

Since ReL pyrocarbon was still unknown through the classical chemical vapour deposition/infiltration (CVD/CVI) conditions, the well-known low temperature textural transition usually referred to as smooth laminar/rough laminar (SL/RL) was in reality a SL/ReL transition. The similar anisotropy of both RL and ReL, as assessed by their respective Ae and OA values, indeed led to a misunderstanding of the pyrocarbon texture/structure. Other recent works are also difficult to consider with the current classification. For instance, the highly textured pyrocarbon mentioned by Reznik and Hüttinger [5] can be either considered as ReL or RL. Thus, the correlation between the growth mechanisms and the CVD/CVI conditions of deposition of RL are still a subject of controversial debate, essentially owing to inappropriate tools for the characterization of the texture/structure of pyrocarbons.

Describing two distinct features (i.e., texture and structure) with only one parameter (usually Ae) is insufficient to discuss the growth mechanism of pyrocarbons adequately. The aim of the present work is to overcome inconsistencies, which may arise (e.g., the confusion between RL and ReL), by proposing other complementary techniques and new textural/structural parameters for a relevant analysis of pyrocarbons.

Six different C/C composite matrices (with rather different structural and textural features), belonging to the three main types of pyrocarbon: RL, ReL and SL, have been considered for the present study.

The first technique, which has been used for the assessment of the structural properties, is Raman microspectroscopy (RMS). Raman spectra of carbon materials are indeed very sensitive to structural defects. The FWHM value of the different bands of the first order Raman spectra of $sp^2$ carbons has been related for years to the density of in-plane structural defects in carbons [7] and [8].

The second technique, considered for the texture analyses, is photo-spectroscopic microscopy. Different authors developed quantitative automatic [9] or semiautomatic [10] procedures based on polarized microscopy. The present approach is an improvement in the measurement of the extinction angle (Ae) [10]. A microspectrophotometer was used to determine precisely Ae at a given wavelength by a complete fitting of the extinction curves. This procedure also leads to another parameter characteristic of pyrocarbons, the optical phase shift ($\delta$ in nm) between the ordinary and extraordinary beams reflected parallel and perpendicular to the graphene plane. Ordinary and extraordinary reflectance can also be deduced from the same technique and compared to values determined in a more usual way. Accurate values of Ae and optical phase shift will be given and discussed for the three types of pyrocarbon studied.

Thereafter, angular resolved electron energy loss spectroscopy (EELS) of the C-K transitions have been applied to RL, ReL and SL, in order to resolve the anisotropy of their electronic properties. The EELS analysis gives both the in-plane and out-of-plane dielectric components, as well as the hybridization of the carbon atoms in the material.

## 2. Experimental procedures

### 2.1. Description of the materials

Two different industrial pyrocarbons, considered as reference materials were processed according to classical isothermal/isobaric conditions from a mixture of propane and methane. They were found to belong respectively to the rough laminar (RL) and smooth laminar (SL) classes of pyrocarbon [11].

The third type of pyrocarbon, ReL, was processed by pulse-CVI ($T = 1000$ °C, $P = 1.2–2$ kPa) from pure toluene, in a small pilot furnace [2].

Two other specimens, i.e., three other pyrocarbon matrices were deposited by CVD or infiltrated by CVI in fibrous preforms from propane at 950 °C, using different pressures ($P$) and residence times ($t$). More details about the experimental apparatus can be found elsewhere [11]. The first specimen, referred to as P1, was processed at $P = 5$ kPa and $t = 3$ s in CVD conditions. The second specimen was prepared at $P = 0.5$ kPa and $t = 0.065$ s. Under these conditions, a gradient in the properties of the pyrocarbon matrix was found from the surface of the fibrous prefom to the core (due to local changes in the composition of the gas phase through the preform). The pyrocarbon matrix was therefore examined distinctively in both CVD (P3s) and CVI (P3c) conditions.

All the as-processed pyrocarbons (industrial and experimental) were prepared within the same temperature range (950–1000 °C). Some of the specimens were submitted to post-heat treatments before the structural investigations. They were annealed 10 min in argon at temperatures ranging from $T_{HT} = 1600$ to 3000 °C.

### 2.2. Preparation of the specimens

The investigations were made using selected area electronic diffraction (SAED), high resolution 0 0 2-lattice fringes imaging (HRTEM), electron energy loss spectroscopy (EELS), optical microscopy, photo-spectroscopic optical analysis, X-ray diffraction (XRD) and Raman microspectroscopy (RMS).

SAED, HRTEM and EELS analyses require electron-transparent samples. Moreover, the EELS technique considered here can be applied only if the average orientation of the carbon layers is perpendicular to the electron incident beam. Hence, in the particular case of EELS analysis, the composites were milled in water according to the conventional technique used for layered materials and deposited on holey carbon films supported by copper grids. Only the particles laid down flat on the grid (with basal planes perpendicular to the electron beam) were submitted to EELS analysis (Fig. 1a).

Classical thinning techniques of transverse cuts of composites (mechanical thinning and Ar$^+$ ion-milling) were applied for the HRTEM analyses as well as for the measurement of OA.

All the other techniques only required polished transverse cuts of the composites.

## 2.3. X-ray diffraction analysis

The XRD measurements were performed with a Siemens D5000 diffractometer ($\lambda$(Cu-K$_{\alpha1}$)) using a $\theta$–$2\theta$ setting. The spectra were recorded directly from the polished cross-sections of the C/C composites. The structural parameters (peak position and width) were indeed found identical from both cross-sections and ground composite powders. Furthermore, the analysis of mere fibres revealed that the influence of the fibres in the composites on the XRD patterns could be neglected, their contribution to the diffraction signal remaining always much lower than that of the various pyrocarbon matrices. A small amount of silicon powder was spread at the surface of the specimen as an internal standard. The $L_a$ in-plane coherence length was derived from the Scherrer equation applied to the 10(0) diffraction band [12].

## 2.4. Raman microspectroscopy

The RMS analyses were performed from the same transverse cuts of polished specimens as those analyzed by optical microscopy. The analyses were conducted with a Labram HR spectrometer (HORIBA Jobin Yvon) with a 632.8 nm emission line, the incident laser being polarized along the anisotropy planes of the transverse cuts of the pyrocarbon coatings. The laser power was kept below 0.5 mW to avoid heating of the sample. The objective of the microscope ($\times$100) allowed a lateral resolution of about 1 μm and the thickness analysed was in the range of 50–100 nm. The band fitting of the first order spectra was carried out with LabSpec 4.14 (HORIBA Jobin Yvon).

## 2.5. Selected area electron diffraction and HRTEM 0 0 2-lattice fringes imaging

The SAED and 0 0 2-lattice fringes imaging were conducted with a CM30ST electron microscope, at an accelerating voltage of 300 kV. The anisotropy of the various pyrocarbons was assessed through the image analysis of the SAED pattern, by measuring the azimuth angle (OA) of the 0 0 2 diffraction arcs [4]. The area selected was 110 nm in diameter (using the smallest diaphragm available), a value sufficiently low for a high spatial resolution, but large enough to avoid any scale effect on the measurement of the texture (i.e., no significant change of OA was found using larger selected apertures).

## 2.6. Electron energy loss spectroscopy

The experimental set-up used for the EELS studies is described in details by Laffont et al. [13] and [14]. The EELS experiments were performed using a Philips CM30ST microscope operating at 100 kV and equipped with Gatan 666 parallel electron energy-loss spectrometer (PEELS). The resolution was approximately 0.9 eV and the electron probe size about 100 nm.

To resolve the anisotropy of the electronic structure, EELS of the outer shell (K level) were recorded from grains of milled matrix oriented perpendicular to the electron beam, according to the diffraction mode in specific conditions (Fig. 1) [14] and [15]. This approach is based on the work of Browning et al. [15], assuming that the intensity of the EELS spectrum at high energy is a linear combination of the in-plane and out-of-plane dielectric constants. The two proportional factors a and b are both related to the convergence and scattering angles ($\alpha$ and $\beta$, respectively) (Fig. 1b). The incident beam being parallel to the average $c$ axis of the sample, an increase of the collection angle allows a wider collection of the carbon plane signal. Two sets of experimental conditions are needed to separate the parallel and perpendicular components from the signal in the imaging mode: $α = 2.2$ mrad; $β = 3$ mrad and $α = 2.2$ mrad; $β = 8.1$ mrad, respectively.

The calculation of the carbon hybridization is finally achieved through the method described by Browning et al. [15]. The dielectric constants parallel ($ε_∥$) and perpendicular ($ε_⊥$) to the averaged $c$ axis are used to build the theoretical EELS spectrum of a virtual isotropic-like material (i.e., containing carbon planes with a random orientation). At high voltage, the scattered intensity is related to the orientation $χ$ of the $c$ axis to the incident beam by

$$I(χ) = a(χ) \cdot ε_⊥ + b(χ) \cdot ε_∥,$$

where $a(χ)$ and $b(χ)$ are two coefficients defined in Ref. [15]. The averaged isotropic-like spectrum is obtained by integrating the intensity from $χ = -π/2$ to $+π/2$. The virtual spectrum of each pyrocarbon can be compared to the isotropic-like virtual spectrum obtained from a graphite single crystal (100% $sp^2$) using the same process (i.e., with single crystals having random orientation). The $sp^2$ content of the pyrocarbon is subsequently calculated from more conventional calculation [15] and [16].

This method is more reliable than the usual technique which simply consists in comparing the two spectra obtained for a pyrocarbon and polycrystalline graphite with a single set of experimental conditions. At the scale investigated by EELS, polycrystalline graphite can not indeed be considered as a randomly oriented carbon planes with a pure $sp^2$ hybridization.

## 2.7. Photo-spectroscopic microscopy

The system used consisted of a NIKON ECLIPSE microscope mounted with a ×100 objective, an infra-red filter, a halogen light source and two rotating polarizer and analyzer. The photo-spectrometer was a PARISS system, with a Pelletier cooled CCD detector. The resolution was about 0.1 nm at the wavelength chosen for the analyses, and the analyzed area was $0.162 × 0.255$ μm$^2$.

A uniform cylindrical layer of pyrocarbon deposited on an isolated fibre is selected from the polished cross-section of the composite. The area of the coating analysed is selected at a place where the growth direction makes an angle of $π/4$ with the incident light polarization (**P**, Fig. 2). The linearly polarized beam is reflected along the two main directions of the material, i.e., parallel, $\mathbf{y}_o$ and perpendicular, $\mathbf{y}_e$ to the basal planes (Fig. 2). The two reflected waves have different amplitude and phase shift. They both interfere after projection ($\mathbf{y}_e → \mathbf{z}_1$, $\mathbf{y}_e → \mathbf{z}_2$) onto the plane of the analyzer, rotated by an angle $θ$ from the crossed polarization position (Fig. 2). The resulting intensity as a function of $θ$, at an angle $π/4$, is thus given by Eq. (2) from Vallerot and Bourrat [10]:

$$I(\theta) = \frac{I_0}{2} \left[ R_e \sin^2\left(\frac{\pi}{4} - \theta\right) + R_o \cos^2\left(\frac{\pi}{4} - \theta\right) \right.$$
$$\left. - 2\sqrt{R_e R_o} \sin\left(\frac{\pi}{4} - \theta\right) \cos\left(\frac{\pi}{4} - \theta\right) \cos\left(\frac{2\pi}{\lambda}\delta\right) \right]$$

where $R_e$ and $R_o$ are the extraordinary and ordinary reflectances, $I_0$, the absolute incident intensity, $\theta$, the rotating angle of the analyzer from the crossed polarization position, $\lambda$, the wavelength used for the measurements and $\delta$, the optical phase shift between the ordinary and extraordinary reflected waves. An accurate value of Ae is therefore obtained after a fit of the complete curve, Ae corresponding to the first minimum of the function $I = f(\theta)$. Other parameters such as $R_e$, $R_o$ and $\delta$ can also be deduced from the extinction curve fit, the absolute intensity $I_0$ being calibrated beforehand. All the curves were recorded at $\lambda = 550$ nm for comparison, since many optical indices of graphitic materials have been measured at this wavelength [17] and [18]. As a matter of fact, optical indices of graphitic materials vary with the wavelength. For instance the calculated extinction angle for graphite, using the optical indices reported by Greenaway et al. [18], is 24° at 450 nm and 26° at 750 nm. The value of the extinction angle for graphite or pyrocarbons will therefore vary whether the light source is a halogen or a xenon lamp, when measured by eyes. This discrepancy has been resolved using a spectrometer mounted on the microscope, allowing the measurement of the extinction curve at 550 nm [10].

## 2.8. Relationship between the matrix areas studied by the different techniques

The spatial resolutions of Raman microspectroscopy and optical microscopy are rather similar and, as specified above, the areas of pyrocarbon analysed with both techniques are approximately the same.

The zones investigated by TEM were chosen within the same matrix regions and the area analysed depends on each specific mode. The probe size for SAED (assessment of OA) and EELS (C-K edge) is about 0.1 μm, i.e., smaller than for RMS (1 μm). However, this value is sufficiently large to prevent any scale effect on the measurement of the texture. The area observed on a 0 0 2-lattice fringes image is obviously less significant (few tenths of nanometers) than for the two other techniques. The HRTEM analysis was conducted from several locations of the thinned sections (within the same matrix parts as those investigated with the other techniques) and a representative image of each specimen has been presented.

XRD provides information on the whole composite. Although the contribution of the fibres could be neglected for the specimens considered, this technique offers no local information on the carbon matrix. XRD was therefore only applied for pyrocarbon matrices exhibiting no marked structural/textural gradient.

# 3. Results and discussion

## 3.1. Qualitative optical microscopy

Within the first group of highly anisotropic pyrocarbons, RL and P3c show common features, at least as observed by optical microscopy (Fig. 3a and d). In the cross-polarization configuration, the Maltese-crosses appear rough and domains of several micrometers show a coherent aspect (characterized by a uniform intensity). According to the first classification proposed by Granoff [19], RL and P3c would be categorized as rough laminar pyrocarbons. The two other anisotropic pyrocarbons P1 and ReL exhibit, on the contrary, very regular and well defined Maltese-crosses, as seen on Fig. 3b and e. Such a high anisotropy (characterized by a high reflectance) and regular aspect of the Maltese-cross can be associated with the regenerative laminar type (ReL) [2]. Highly anisotropic pyrocarbons can obviously be divided into two groups: rough and regenerative laminars.

P3s and SL also show a regular Maltese-cross (Fig. 3c and f) but, in contrast, they are characterized by a poor anisotropy (as shown by the low reflectance). Therefore, they can both be assumed to belong to the smooth laminar category.

A qualitative analysis of pyrocarbons by polarized/analyzed light microscopy clearly shows that the highly anisotropic class can actually be divided into two groups. This distinction agrees the previous results of Bourrat et al. [2] and is of great importance as it may give rise to very different physical properties (e.g., the respective Young's modulus of RL and ReL [20]). This feature points out the need of a new quantitative and unambiguous structural parameter, as will be discussed below.

## 3.2. Quantitative photo-spectroscopic microscopy

The normalized extinction curves ($I(\theta)/I_0$) of the various pyrocarbon materials were acquired [10]. The optical parameters, i.e., the extinction angle Ae, the phase shift $\delta$ and the ordinary and extraordinary reflectances $R_o$ and $R_e$ have been determined from the fit of the $I(\theta)/I_0$ functions. The values are listed in Table 1 together with the corresponding extinction angle determined visually by the usual technique (Ae′) and the parameter OA measured by SAED.

The extinction angles measured at 550 nm are consistently lower than those assessed visually (Table 1). The calculated extinction angles increase with the wavelength because of the variation of the optical indices of graphite within the visible range [10] and [21]. The optical phase shifts obtained from the extinction curves agree relatively well with the theoretical values for graphite at 550 nm (i.e., 25 nm, when considering the optical indices determined by Greenaway et al. [18] or 41 nm, using those reported by Ergun et al. [17]).

The Ae and OA values show two main groups of data corresponding to two major types of pyrocarbons. RL, ReL, P1 and P3c exhibit high Ae and low OA values, which can be associated to anisotropic pyrocarbons, whereas SL and P3s show lower Ae and higher OA values. The reflection ratio $R_e/R_o$ varies approximately linearly with the anisotropy, as measured by the extinction angle (Fig. 4). The optical phase shift ($\delta$) is also found to be closely related to the anisotropy as measured by electronic diffraction (Table 1). Some discrepancies clearly appear between the anisotropy degrees respectively assessed by optical (Ae) and electronic (OA) microscopies.

Nevertheless, the comparison between the optical results and the OA values indicate that the optical properties of pyrocarbons are closely related to their textural properties. This feature has already been pointed out by Bourrat et al. [4]. It will be developed below through the use of angular resolved EELS. On the basis of their OA values only, RL, ReL, P1 and P3c would be categorized as a highly anisotropic pyrocarbons, whereas P3s and SL would be rather regarded as medium anisotropic.

### 3.3. Raman microspectroscopy

As other graphite-like materials, the first order Raman spectra of pyrocarbons exhibit the following main features: the in-plane mode with the $E_{2g}$ symmetry, first identified from a graphite single crystal at a frequency of 1575 cm$^{-1}$ by Tuinstra and Koenig[2] [22], the disorder induced D band [22], observed at ≈1330 cm$^{-1}$ for a laser excitation of 1.97 eV (wavelength 632.8 nm) and the D″ band, attributed to an amorphous form of sp$^2$ carbon, at about 1500 cm$^{-1}$ [23].

When pyrocarbons are heat treated, a D′ band, also disorder induced, clearly appears around 1620 cm$^{-1}$. Moreover, if the incident light is polarized perpendicularly to the planes, a sharp "A" band is observed at 867 cm$^{-1}$ [24]. This feature has been assigned to a particular vibrational mode in graphite, with "out-of-plane" atomic displacements [25]. It is also visible for as-deposited pyrocarbons, but too weak and too broad to be accurately and systematically characterized within this study [24].

For a few years, the dispersive effect of the D mode has been explained by resonant phenomena occurring between electron transitions and phonons having the same wave vector near the K point of the Brillouin zone [26]; this was known as the **k** = **q** quasi-selection rule. This dispersive effect is now rather attributed to a double resonance Raman scattering mechanism (DRRS) [27]. Most of the features in the spectra of sp$^2$ carbonaceous materials (e.g., the D, D′ and D″ bands) can be predicted by means of the DRRS theory [28]. The phonon modes experimentally observed for a laser energy $E_L < 3$ eV can be successfully correlated with the phonon dispersion curves, covering a large area of the 2D graphite Brillouin zone [29].

The contribution of the 1620 cm$^{-1}$ D′ component to the main broad Raman feature lying in the G band region (1500–1700 cm$^{-1}$) in the spectra of pyrocarbon matrices (either as-processed or treated at low temperatures), was established by Vallerot et al. [24] by comparing the 632.8 nm-excitation spectra with those obtained with a UV-excitation (325 nm). This study shows that a classical single-peak deconvolution of the main 1500–1700 cm$^{-1}$ feature (i.e., ignoring the D′ component) may lead to significant inaccuracies in the frequency and the width of the $E_{2g2}$ mode. This may arise especially for as-deposited materials, where both the G ($E_{2g2}$ mode) and D′ bands are particularly broad and overlapped. Experimental results from graphitization studies of carbon materials showed that the FWHM of both the D and G bands are correlated with the structural state of the material [8]. A recent systematic study of a large number of as-deposited pyrocarbons has shown that when a five band fitting procedure is applied to the various first order spectra, a linear law is established between the FWHMs of the D, D′ and G bands [21]. Therefore, as the D band is usually very strong and easy to fit in the case of pyrocarbons, the width of the D band (FWHM$_D$) was chosen as the in-plane structural ordering parameter.

All the first order Raman spectra were fitted with Lorentzian functions for the D, G and D′ bands and a Gaussian for the I and D″ bands (Fig. 5). On the basis of the analysis of numerous spectra obtained from various pyrocarbons, either heat-treated or not, and with both visible and UV excitations, the frequencies of certain minor peaks were kept constant ($v_I = 1170$ cm$^{-1}$, $v_{D'} = 1618$ cm$^{-1}$ and $v_{D''} = 1500$ cm$^{-1}$) during the fitting procedure, to limit the uncertainty in the spectral parameters. Moreover, the parameters of the D band (e.g., FWHM$_D$, of particular interest in this study) were found, in all cases, to be only little influenced by the other Raman peaks during the fitting procedure.

The Raman features of the various pyrocarbons are obviously different. The bandwidth, in particular, increases considerably from RL to ReL (Fig. 5). The structural parameter which has been proposed above, FWHM$_D$, is very sensitive to the low energy defects. It indeed strongly decreases to a common value for all pyrocarbons after a heat treatment at 1600 °C (Fig. 6). FWHM$_D$ subsequently decreases linearly up to 2000 °C and remains almost constant beyond this temperature. These structural defects, healed to a large extent after a heat treatment at 2000 °C, were assumed to be in-plane local disorientations or dislocations by Rouzaud et al. [30], in their study of pyrolytic carbon films.

Beyond 2000 °C, most of the local disorientations have disappeared to form straight graphene layers. This is confirmed by the 0 0 2-lattice fringes micrographs of the heat-treated LR and ReL pyrocarbons. The carbon layers of both pyrocarbons have indeed clearly straighten out from $T_{HT} = 1800$ °C to 2200 °C (Figs. 7b, c and 8b, c). From $T_{HT} = 2000$ °C to 3000 °C, the pyrocarbons are submitted to the ultimate stage of graphitization, corresponding to a lateral extension of the graphene planes (Figs. 7c and 8c) [30]. This two-stage structural improvement is accurately monitored by means of $L_a$, the in-plane coherence length determined by XRD (Fig. 9). Conversely, the graphene planes do not expand significantly during the early stages of graphitization, $L_a$ remaining almost constant up to 2000 °C [30].

The two main stages of the graphitization process can be evidenced in Fig. 9, as the heat treatment temperature increases. The first stage is characterized by a large decrease of FWHM$_D$ up to 2000 °C and corresponds to the gradual straightening of the graphene planes (Figs. 7a, b and 8a, b). The second stage ($T_{HT} > 2000$ °C) is defined by the lateral extension of the layers (Figs. 7c and 8c) and results in the increase of $L_a$ (Fig. 9), while FWHM$_D$ remains almost equivalent for all the pyrocarbons (Fig. 6 and Fig. 9). The latter parameter therefore appears as a very reliable indication of the heat treatment temperature encountered by pyrocarbons (at least below $T_{HT} = 2000$ °C).

The sensitivity of FWHM$_D$ to the in-plane defects makes this parameter particularly relevant for the structural characterization of as-deposited pyrocarbons. FWHM$_D$ is plotted as a function of the corresponding OA value in Fig. 10, for various as-processed pyrocarbons. The FWHM$_D$ values vary significantly with the type of pyrocarbons. The high FWHM$_D$ value of ReL indicates that its local structure is likely much more disordered (by disorientations and/or in-plane defects) than that of RL. This type of defects can be partially evidenced by the HRTEM analysis of the 0 0 2-lattice fringes (Fig. 11). The ReL pyrocarbon exhibits highly curved graphene layers (Fig. 11a) which result in particularly broad Raman features and a high FWHM$_D$ value (Fig. 10). Conversely, the less curved layers observed successively for the SL and RL (Fig. 11b and c) pyrocarbons (at the HRTEM scale), give rise to consecutively lower FWHM$_D$ values (Fig. 10). It is worthy of note that this structural parameter derived from RMS is totally independent of the anisotropy of the material. The diagram presented in

Fig. 10 clearly distinguishes the structure of pyrocarbons having an equivalent anisotropy. The anisotropy parameter (OA) alone, is clearly insufficient to thoroughly classify pyrocarbons. A two-dimension diagram is therefore proposed, showing both the textural (OA) and structural (FWHM$_D$) parameters as measured by TEM and RMS respectively (Fig. 10). From the measurement of these two parameters for a larger variety of specimens [21], a systematic classification of pyrocarbons can be proposed, evidencing well defined textural/structural domains (Table 2).

Continuous transitions can be observed between the various domains, suggesting that series of pyrocarbons were processed according to a gradual change of the processing conditions. Rough laminar (high anisotropy, low amount of structural defects) and Regenerative laminar (high anisotropy, large amount of structural defects) are now clearly differentiated.

Moreover, by simply comparing the samples processed in the laboratory (i.e., in well defined conditions) with the industrial reference materials, it can be postulated that the two anisotropic pyrocarbons, ReL and RL were prepared under very different conditions. More details on the synthesis of such pyrocarbons are given by Vignoles et al. [31]. Le Poche et al. [6] have shown than Regenerative laminar pyrocarbon is deposited when the maturation of the gazeous phase is high enough to produce large planar aromatic species, which are deposited by physisorption onto the substrate prior to form pyrocarbon. Conversely, rough laminar pyrocarbon is deposited by the chemisorption of smaller species produced from the very early stages of the propane decomposition. When reactive sites are present in sufficient number to allow the chemisorption of the most reactive species, i.e., with high surface/volume ratio (e.g., within a fibre preform in CVI conditions), RL pyrocarbon is deposited. On the other hand, under the same processing conditions, but at the surface of the felt (in CVD conditions), where the surface/volume ratio is much lower, SL is deposited. A comprehensive modelling of the growth mechanism of pyrocarbon is proposed in Ref. [31].

### 3.4. Angular resolved EELS

3.4.1. Dielectric components parallel and perpendicular to the graphene plane

The deconvolution of the EELS signal collected allows to separate the contribution perpendicular to the main local plane axis, $\varepsilon_\perp$, from the component parallel to the local axis, $\varepsilon_\parallel$. The reference material used was a natural crystal of graphite from Sri Lanka. The spectra are normalized with regard to the maximum intensity of the main feature at about 292 eV. The angular resolved EELS analyses being made on the C-K transition, the Fermi level is considered to be equal to about 278 eV (the energy needed to excite the electron out of the K level).

The absence of the 1s → π* component (at 285 eV) on the various $\varepsilon_\perp$ spectra confirms the accuracy of the deconvolution. It also indicates that the dielectric constants along the planes of the pyrocarbons are almost equivalent to that of graphite (Fig. 12b). This feature is indeed consistent with the assumption of the uniaxial crystal (such as graphite), which is required for the calculation of the $\varepsilon_\parallel$ and $\varepsilon_\perp$, components [15]. The spectrum of the $\varepsilon_\perp$ component shows a sharp transition 1s → σ* around 290 eV and the intensity remains almost constant at a higher energy (Fig. 12b). The spectrum of the $\varepsilon_\parallel$ component generally shows a transition 1s → π* characterized by a sharp and intense peak (Fig. 12a). However, some slight differences between the various pyrocarbons and the graphite specimen are observed at 285 eV. The decrease of the 285 eV peak intensity is supposed to be related to the sp$^{2+\varepsilon}$ hybridization of the material (see Section 3.4.2). Beyond 292 eV, the signal can be considered as the density of states above the Fermi level. Only few features are visible in this part of the spectra of the pyrocarbons. Their structural defects might indeed be responsible for a larger number of allowed electronic transitions, as compared to ideal graphite.

However, the electronic properties of pyrocarbons, especially parallel to the graphene planes were found to be similar to those of the graphite reference. This result confirms the assumptions stated above, presuming that the texture (i.e., the anisotropy) was the most influential factor for the optical properties of the various pyrocarbons.

3.4.2. Hybridization of pyrocarbons

In their study of the EELS C-K edge of pyrocarbons, Reznik et al. [32] presented paradoxical results concerning SL and RL pyrocarbons. SL was indeed found to contain more than 100% $sp^2$ carbon atoms, whereas RL was 100% $sp^2$ hybridized. However, it is worthy of note that the orientation of the anisotropic texture was not taken into account in the above study. The use of angular resolved EELS might have led to more consistent results. In the technique developed by Browning et al. [15], each $\varepsilon_\parallel$ and $\varepsilon_\perp$ component determined from various pyrocarbons and graphite were used to calculate an averaged spectrum over all the directions, for the various pyrocarbons and graphite. The resulting spectra can be assumed to be typical of isotropic-like materials.

The spectra are represented in Fig. 13, after their intensity being normalized to the main feature at 292 eV (characteristic of σ transitions). The intensity decrease of the 285 eV component is therefore directly proportional to the deficit of $sp^2$ hybridized carbon atoms. Assuming ratios of four σ bonds to zero π in a purely $sp^3$ hybridized carbon (diamond) and three σ to one π in a 100% $sp^2$ (graphite), the $sp^2$ hybridization percentage of the pyrocarbons can be calculated by the method proposed by Lossy et al. [16] using a 284–287 eV interval for the integration of the intensities. Results are given in Table 3.

The question which may raise from these data, concerns the nature of the non-$sp^2$ hybridized carbon atoms, which are present in almost the same proportion in the three different pyrocarbons. The Raman analyses do not support the occurrence of a significant amount of purely $sp^3$ (diamond like) C–C bonds (sharp single peak at 1332 $cm^{-1}$, corresponding to the **q** = **0** triply degenerate optical phonon) in pyrocarbons. Furthermore, an amount as high as 20% of $sp^3$ bond in the carbon material associated with a low hydrogen content (H/(C + H) ratio < 7 at.%) [2] and a low nanoporosity, would give rise to a particularly high density (beyond the value of graphite), in contradiction with the microscale densities measured experimentally (densities of 1.95, 2.11 and 2.13 g $cm^{-3}$ were reported for SL, ReL and RL pyrocarbons, respectively [2]). Therefore the deficit of purely $sp^2$ carbon atoms can be more precisely related to an intermediate $sp^{2+\varepsilon}$ hybridization. This type of hybridization has already been mentioned by Hiura et al. [33] in the case of rippled graphene planes. These authors proposed seven examples of $sp^3$-like in-plane defects, propagating along the 1 0 0 and 2 1 0 symmetry axes.

These defects give rise to a rippled aspect (with a 5–20 nm period) of the graphene plane which propagates at large scale along the symmetry axes. These line defects might be responsible for the distortions of the graphene planes in pyrocarbons, which are easily noticed by HRTEM. The presence of pentagons and heptagons might also induce an average $sp^{2+\varepsilon}$ hybridization. The occurrence of $C_5$ and $C_7$ rings within the graphene layers have already been proposed in the case of a low anisotropic matrix such as SL [34]. Furthermore, pentagons could explain the lower value of the $sp^2$ ratio of SL, as compared to that of the RL or ReL pyrocarbons.

## 4. Conclusions

The $FWHM_D$ parameter has been shown to be sensitive to the low energy structural defects in pyrocarbons, i.e., defects eliminated by a 2000 °C heat treatment. These defects might be associated to disorientations of the graphene layers and possibly also to in-plane structural defects such as local dislocations. Besides, the anisotropy of the graphene layers has been quantified by electronic diffraction (OA, as measured from the SAED patterns).

A new classification of pyrocarbons is proposed, consisting in a two dimension diagram displaying both the structural ($FWHM_D$) and textural (OA) parameters. This approach has allowed to differentiate pyrocarbons deposited under a wide range of processing conditions. In particular, regenerative and rough laminar pyrocarbons could be for the first time quantitatively distinguished.

The optical properties, i.e., the extinction angle, the optical phase shift and the ordinary and extraordinary reflectance, have been accurately determined at 550 nm by means of the extinction curve method. These results have been completed by the in and out-of-plane dielectric constants assessed by angular resolved EELS. These investigations conclude that most of the optical properties are mainly dependent on the texture of pyrocarbons.

Moreover, the hybridization degree of the carbon atoms were assessed by angular resolved EELS. About 80% of the carbon atoms in the pyrocarbons are $sp^2$ hybridized. The lack of pure $sp^2$ carbon atoms, as compared to graphite, is explained by the presence of $sp^3$-like defects. The rippling of the graphene planes might be associated to the $sp^3$-like line defects propagating through the 1 0 0 and 2 1 0 axes of the graphene planes. A particular low amount of the $sp^2$ bonding is found for SL, which might be explained by the presence of pentagons in the matrix.

## Acknowledgements

The authors would like to thank M. Couzi from LPCM, Talence and P. Delhaes from CRPP, Pessac, for fruitful discussions. Prof. P. Touzain is acknowledged for providing the specimens of natural graphite. The authors are also greatly indebted to Snecma and CNRS for providing grants to JMV and AM. Finally, the Conseil Régional d'Aquitaine is acknowledged for its financial support for the purchase of the Raman and photo-spectroscopic equipments.

# Figures

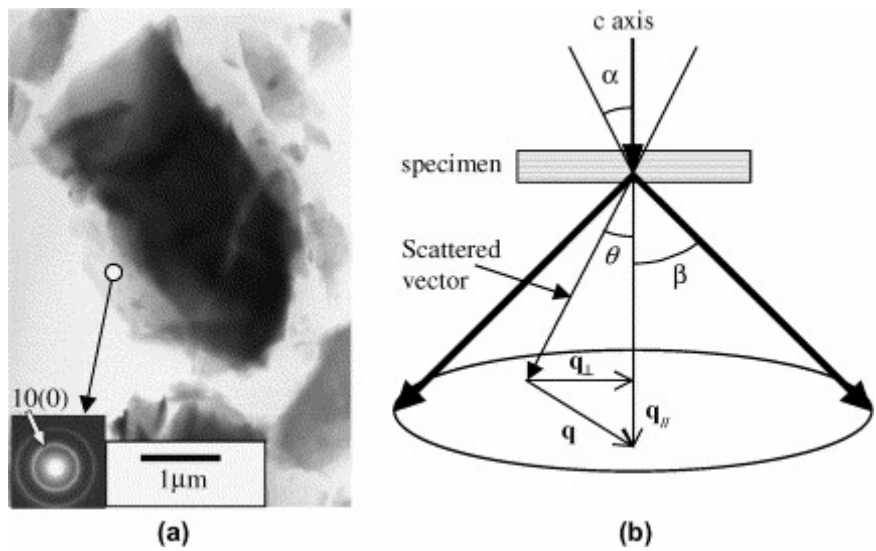

Fig. 1. Angular resolved EELS geometry ((b) see Ref.12).

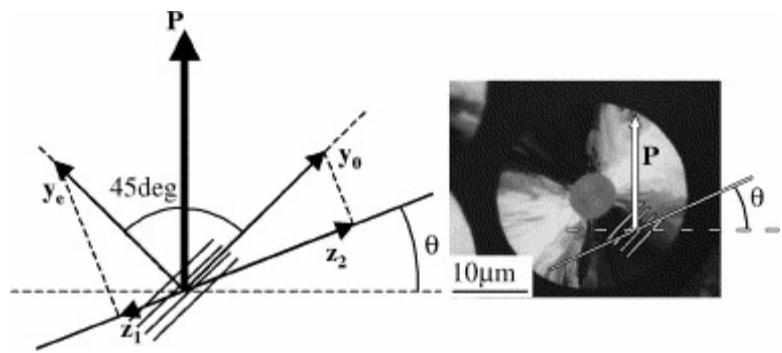

Fig. 2. Optical measurement of the anisotropy (Ae). The incident polarized wave **P** is reflected in two waves along the two main directions of graphite. These two waves interfere onto the plane of the analyzer when rotated to an angle $\theta$ from the $\theta = 0$ crossed position.

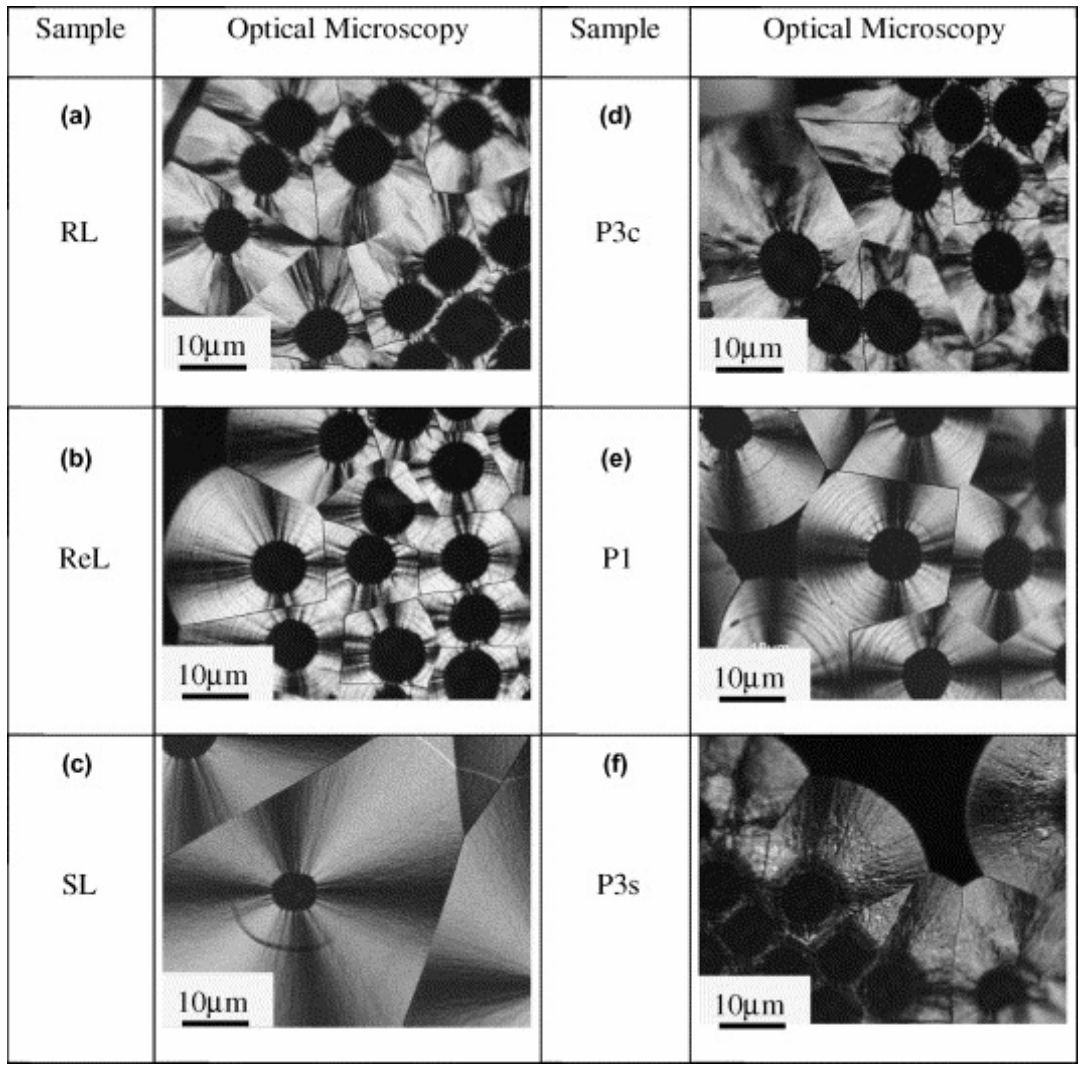

Fig. 3. Comparative optical features of the various pyrocarbons.

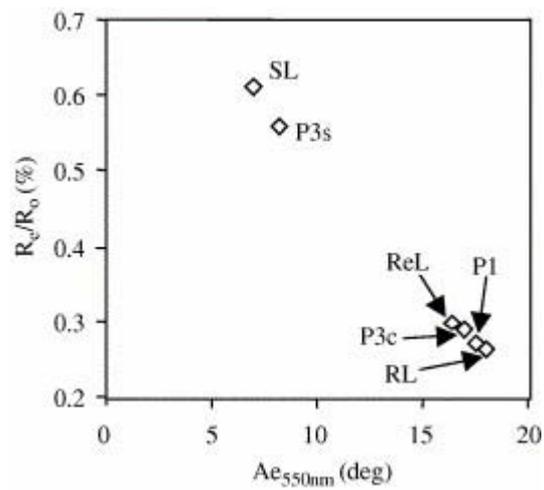

Fig. 4. Relationships between Ae measured at 550 nm and $R_e/R_o$.

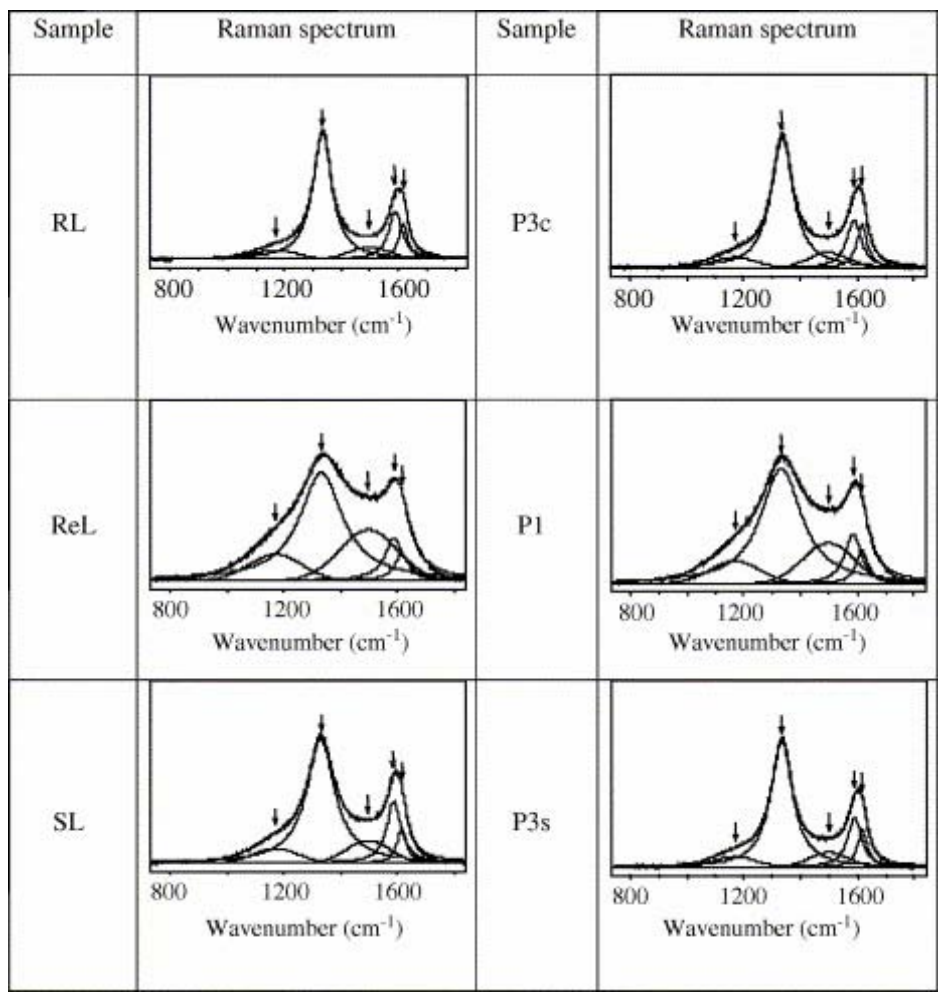

Fig. 5. First order Raman spectra of the various pyrocarbons. Arrows point out from left to right respectively the I, D, D″, G and D′ bands.

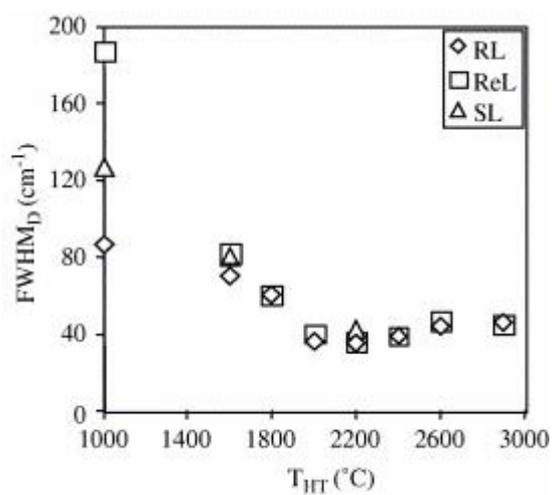

Fig. 6. FWHM$_D$ of the various pyrocarbons as a function of the heat treatment temperature $T_{HT}$.

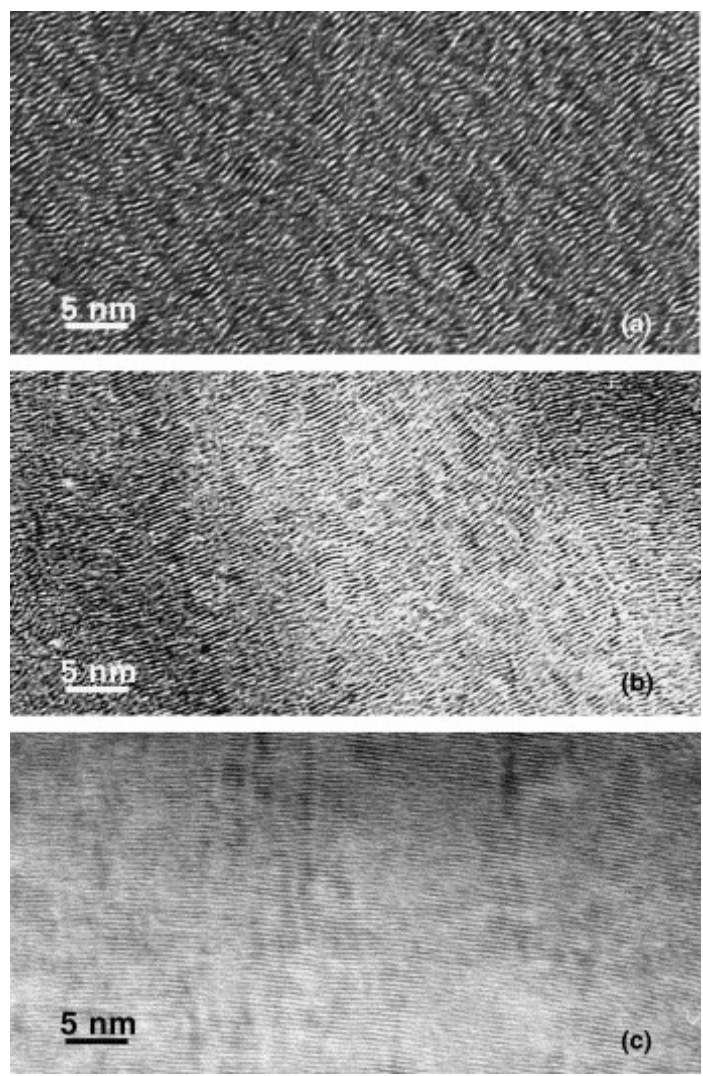

Fig. 7. HRTEM 0 0 2-lattice fringes micrograph of the ReL pyrocarbon: (a) as-processed, (b) $T_{HT} = 1800$ °C and (c) $T_{HT} = 2200$ °C.

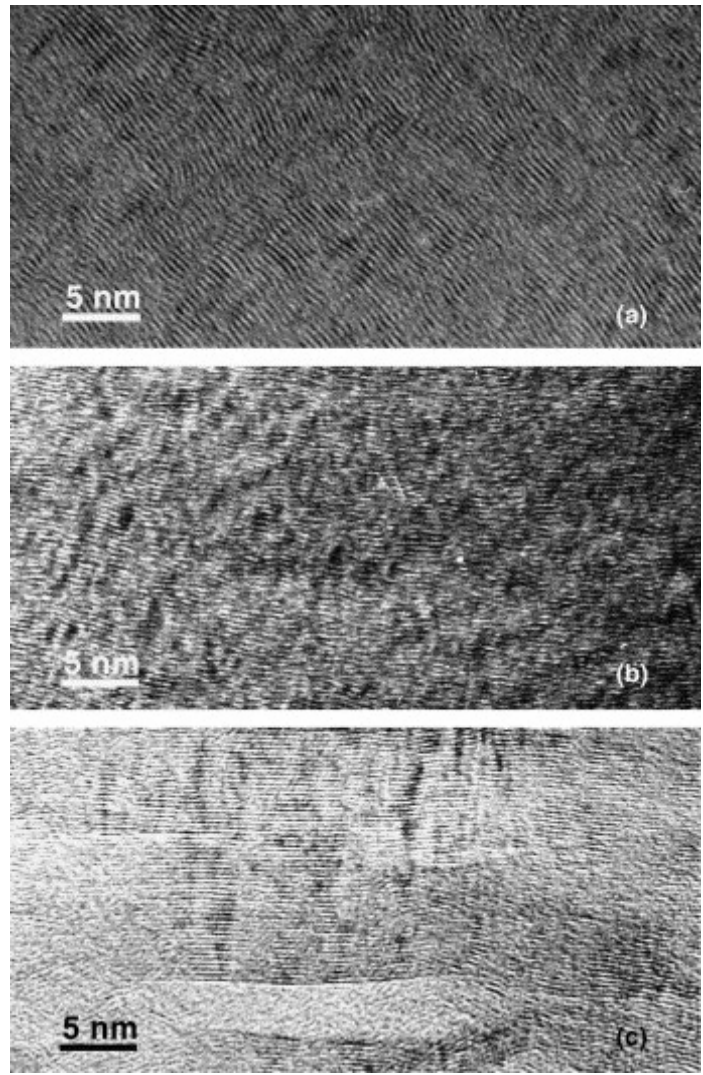

Fig. 8. HRTEM 0 0 2-lattice fringes micrograph of the RL pyrocarbon: (a) as-processed, (b) $T_{HT}$ = 1800 °C and (c) $T_{HT}$ = 2200 °C.

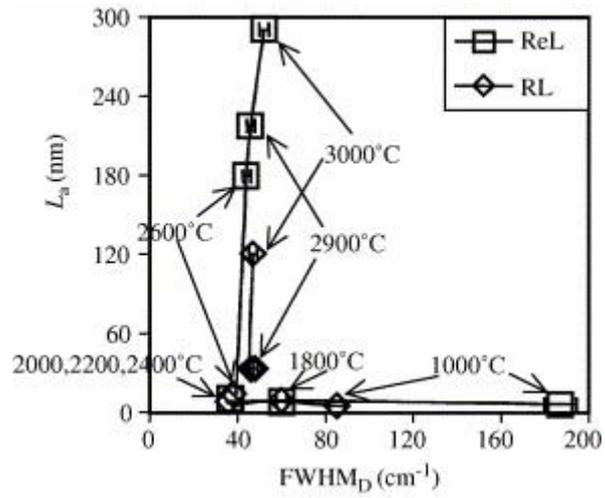

Fig. 9. FWHM$_D$ of the ReL and RL pyrocarbons as a function of $L_a$ and the heat treatment temperature $T_{HT}$.

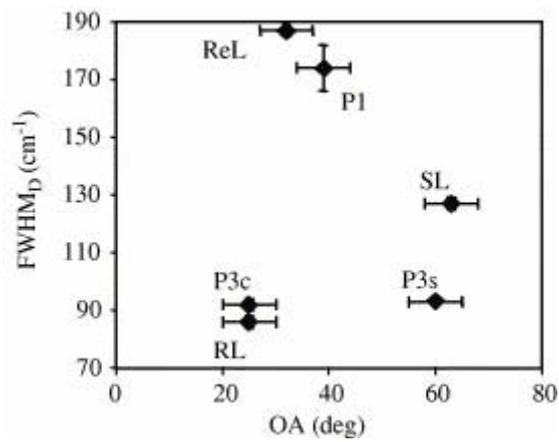

Fig. 10. FWHM$_D$ of the various pyrocarbons as a function of OA.

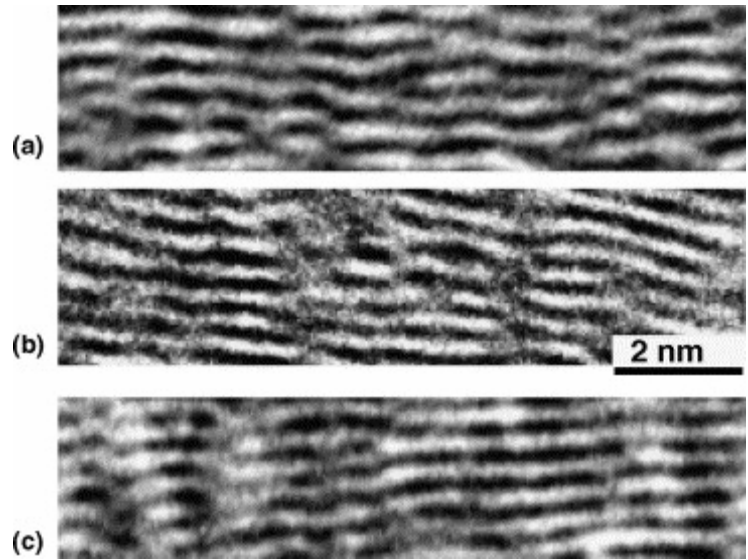

Fig. 11. HRTEM 0 0 2-lattice fringes micrograph of the as-processed pyrocarbons: (a) ReL, (b) SL and (c) RL.

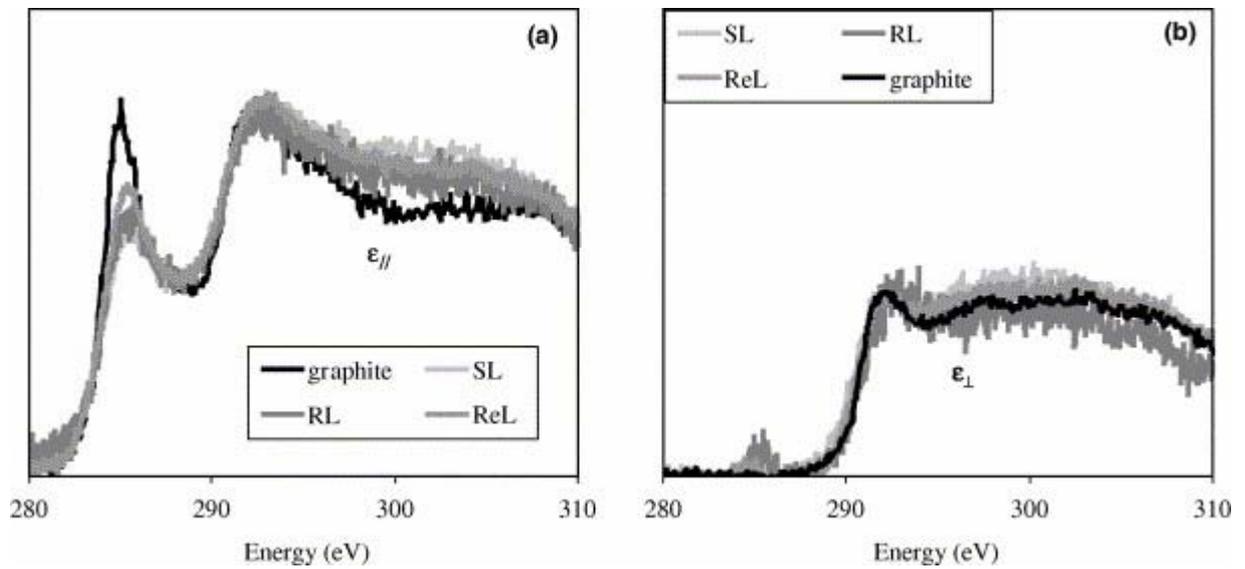

Fig. 12. Dielectric constants parallel (a) and perpendicular (b) to the main local *c* axis. RL, ReL and SL pyrocarbons are compared to natural graphite from Sri Lanka.

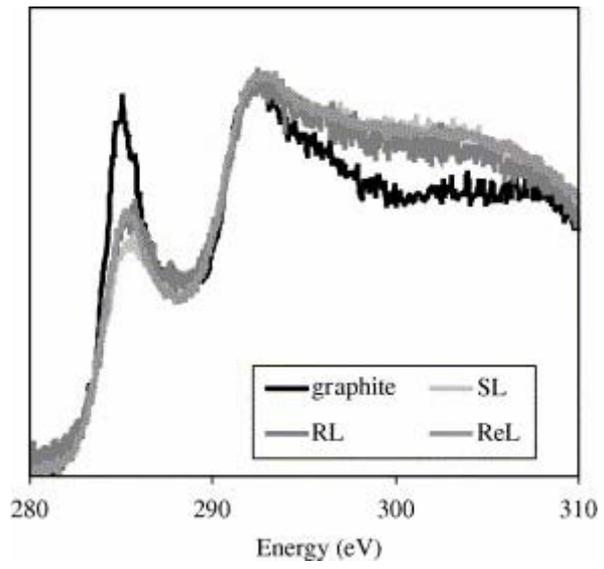

Fig. 13. Calculated averaged spectra of graphite and pyrocarbons.

# Tables

Table 1. : Ae, $\delta$, $R_o$ and $R_e$, deduced from the extinction curves

| Sample | Ae′ (deg) | Ae (deg) | $\delta$ (nm) | $R_o$ (%) | $R_e$ (%) | OA (deg) |
|---|---|---|---|---|---|---|
| RL | 19–20 | 18.0 | 16 | 22.5 | 5.9 | 25 |
| ReL | 19–20 | 16.4 | 7 | 31.9 | 9.5 | 32 |
| SL | 10–11 | 7.0 | 8 | 19.7 | 12.0 | 63 |
| P3c | 19–20 | 17.0 | 18 | 22.8 | 6.6 | 25 |
| P3s | 10–11 | 8.2 | 4 | 18.8 | 10.5 | 60 |
| P1 | 19–20 | 17.5 | 8 | 32.1 | 8.7 | 39 |

Ae′ is the visually measured extinction angle, and OA the orientation angle determined by SAED with a 100 nm aperture.

Table 2. : Classification of pyrocarbons showing various textural/structural domains

| | Rough laminar | Regenerative laminar | Smooth laminar |
|---|---|---|---|
| $FWHM_D$ (cm$^{-1}$) | [80; 90] | [170; 200] | [90; 130] |
| OA (deg) | [20; 30] | [30; 40] | [60; 70] |

Table 3. : Ratio of sp$^2$ to total (sp$^2$ + sp$^3$) bonding in the main types of pyrocarbon, from angular-resolved EELS spectroscopy

| | Rough laminar | Regenerative laminar | Smooth laminar |
|---|---|---|---|
| sp$^2$/(sp$^2$ + sp$^3$) ratios (%) | 82.5 | 83.3 | 79.5 |